\documentclass[aps,12pt]{revtex4-1}
\usepackage{amsmath,amssymb,graphicx,hyperref,float}
\usepackage[small,flushleft,indent]{caption}

\preprint{}

\begin{document}
	
\title{Does the shape of the shadow of a black hole depend on motional status of an observer?}

\author{Zhe Chang}
\author{Qing-Hua Zhu}
\email{zhuqh@ihep.ac.cn}

\affiliation{Institute of High Energy Physics, Chinese Academy of Sciences, Beijing 100049, China}
\affiliation{University of Chinese Academy of Sciences, Beijing 100049, China}


\begin{abstract}
  In a recent work on rotating black hole shadows [Phys. Rev. D{\bf 101}, 084029 (2020)],  we proposed a new approach for calculating size and shape of the shadows in terms of astrometrical observables with respect to finite-distance observers. In this paper, we introduce a distortion parameter for the shadow shapes and discuss the appearance of the shadows of static spherical black holes and Kerr black holes  in a uniform framework. We show that the shape of the shadow of a spherical black hole is circular in the view of arbitrary observers, and the size of the shadows tends to be shrunk in the view of a moving observer. The diameter of the shadows is contracted even in the direction perpendicular to the observers' motion. This seems not to be understood as length contraction effect in special relativity. The shape of Kerr black holes is dependent on motional status of observers located at finite distance. In spite of this, it is found that there is not a surrounding observer who could view the shape of the Kerr black hole shadows as circularity. These results could be helpful for observation of the Sagittarius A* in the centre of the Milky Way, as our solar system is moving around the centre black hole.  
\end{abstract}

{\maketitle}

\section{Introduction}

It is well known that the shadow of a black hole could be used as a tool to seek evidence for exotic matter \cite{davoudiasl_ultralight_2019} or test specific gravity theory \cite{ayzenberg_black_2018,moffat_masses_2020}, Lorentz violence \cite{ding_exact_2020} and so on. Since the first image of a black hole was taken by the Event Horizon Telescope \cite{collaboration_first_2019}, there has been an increase in research of the black hole shadows.

The black hole shadows were first proposed by Synge \cite{synge_escape_1966} in Schwarzschild space-time. For rotating black hole shadows, Bardeen first found that the spin of a black hole would result in distortion of the shape of the shadows \cite{bardeen_timelike_1973}. On the one hand, the studies about size of shadows usually focused on the spherical black holes. One can consider the black hole shadows in the expansion of the universe  \cite{bisnovatyi-kogan_shadow_2018,perlick_black_2018} and coupled to or surrounded by given matter fields \cite{tian_testing_2019,allahyari_magnetically_2020,reji_gravitational_2020,wang_shadow_2020}. On the other hand, the studies about shape of the shadows usually involved rotating black holes. This is because the shapes of the shadows are closely related to parameters of the rotating black holes. As these parameters could be originated from rotating regular black holes  \cite{abdujabbarov_shadow_2016}, various kinds of parametrized Kerr-like black holes \cite{atamurotov_shadow_2013,johannsen_photon_2013} or low-energy limit of string theory \cite{hioki_measurement_2009}, modified gravity \cite{ayzenberg_black_2018,contreras_black_2020,jusufi_rotating_2020}, given matter field   \cite{wei_observing_2013,cunha_shadows_2015,ovgun_shadow_2018,jusufi_black_2019,konoplya_shadow_2019,kumar_shadow_2019,ding_exact_2020,haroon_shadow_2020} of a rotating black hole, one might expect the parameters of a black hole can be constrained from observation of the shadows.

Compared with the usages of the black hole shadows as an observable in various situations, there seems less progress in analytic approaches for calculating the shadows, especially in the case of rotating black hole shadows with respect to finite-distance observers. In the 1970s, Bardeen first calculated Kerr black hole shadow by making use of orthogonal tetrads adapted to zero angular momentum observers (ZAMOs) \cite{bardeen_timelike_1973}. This approach was still used in recent works \cite{johannsen_photon_2013,stuchlik_light_2018}. Also, using orthogonal tetrads adapted to Carter's observers, Grenzebach $et~al.$ considered the Kerr-like black hole shadows for finite-distance observers \cite{grenzebach_photon_2014}. This approach was also used in  Refs.~\cite{mars_fingerprints_2017,eiroa_shadow_2018,haroon_shadow_2019,neves_upper_2020}. In these approaches, a set of orthogonal tetrads should be adopted so that the shadow of a black hole can be calculated at local. Alternatively, there is another approach for calculating the shadow of a black hole without using orthogonal tetrads \cite{chang_revisiting_2020}. The information about appearance of the shadows can be expressed in terms of astrometrical observables with respect to given observers.
For the orthogonal tetrads approach and the astrometrical observable approach, it is, however, not clear whether they are consistent with each other. This motivates us to seek  a more careful comparison.

In addition, as also suggested in Ref.~\cite{chang_revisiting_2020}, the shape of a rotating black hole is dependent on  the observers' motional status. Thus, in this paper, we would further verify this suggestion in a more rigorous manner, and then study whether the shapes of the shadows are different when using different approaches we mentioned above.

This paper is organized as follows. In section \ref{II}, using the astrometrical observables, we introduce a distortion parameter that can quantify the distortion of the shape of the shadows from circularity. In section \ref{III}, by making use of the distortion parameters, we prove that the shape of static spherical black hole shadows is independent on motional status of observers. Then we turn to studying the size of the spherical black hole shadows for different observers. In section \ref{IV}, we show that the shape of the shadows of Kerr black holes is highly dependent on the motion status of observers. We find that the shapes of the shadows calculated with orthogonal tetrad approaches and astrometrical observable approaches are different in the strong gravitational field regime. In section \ref{V}, conclusions and discussions are summarized.

\section{ Distortion parameters of black hole shadows from
circularity}\label{II}

In this section, we briefly review the approach of calculating black hole shadows using astrometrical observables. Then, we introduce a parameter that can quantify the distortion of the shadows deviated from circularity. Here, we consider the most interesting case that observers located at the equatorial plane.

\begin{figure}[ht]
	\centering
	{\includegraphics[width=0.7\linewidth]{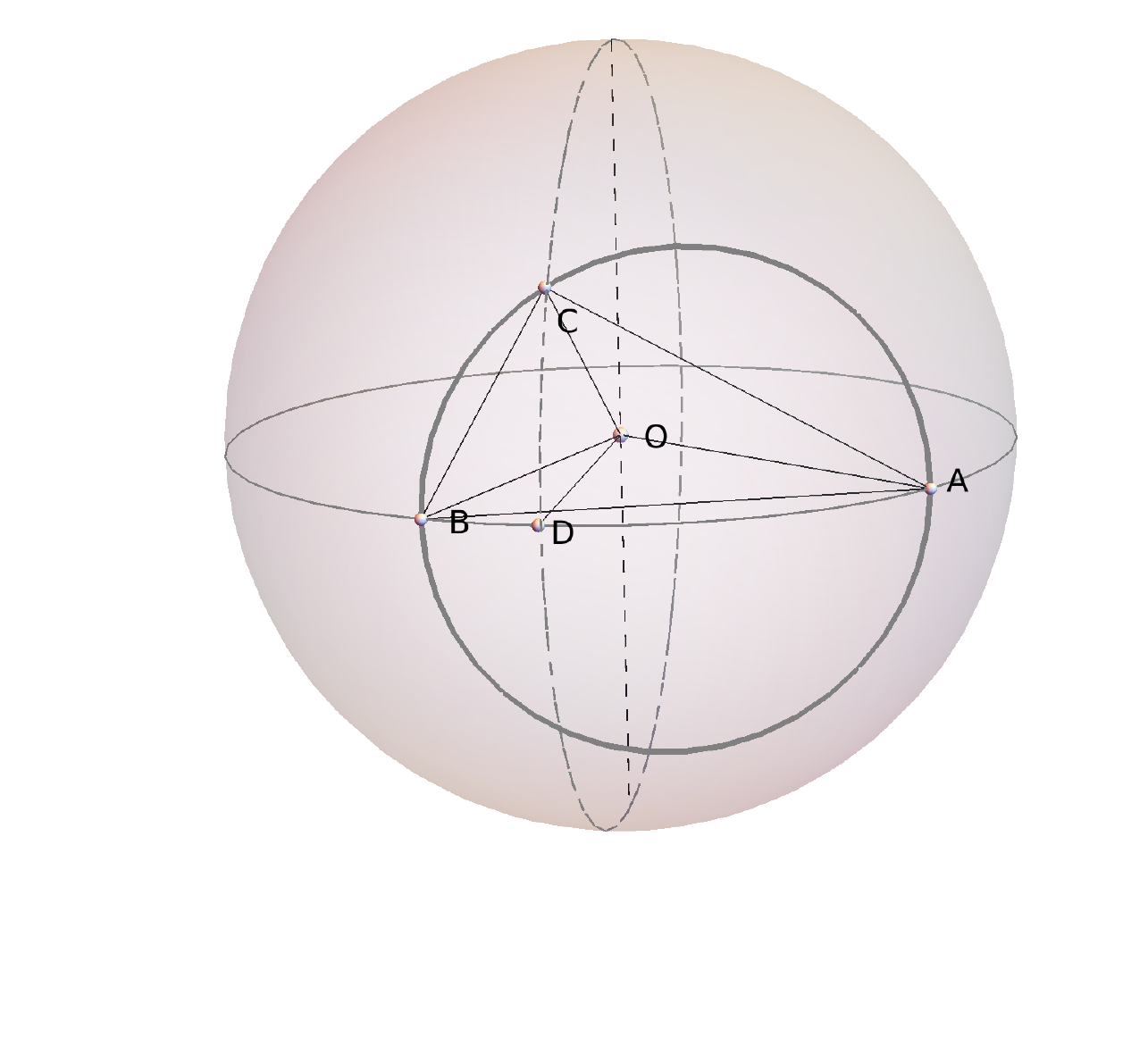}}
	\caption{Schematic diagram of black hole shadows in terms of astrometrical observables in the celestial sphere, where the angles between light rays from photon region are $\alpha \equiv \angle  \rm{BOC}$, $\beta \equiv \angle	\rm{AOC}$ and $\gamma \equiv    \angle \rm{AOB}$, and the celestial coordinates of point $\rm C$ are $\Psi \equiv     \pi/2 -\angle	\rm{COD}$ and $\Phi \equiv    \angle \rm{BOD}$. \label{Fig1}}
\end{figure}

As shown in Ref.~\cite{chang_revisiting_2020}, the size and shape of the shadow of a black hole can be expressed in terms of astrometrical observables $(\alpha, \beta, \gamma)$ in  the celestial sphere. Namely, for the light rays $k$, $w$ and $l$ from photon region of a black hole, the observables in astrometry are the angles between these light rays,
\begin{eqnarray}
  \cos \gamma & = & \frac{k \cdot w}{(u \cdot k) (u \cdot w)} + 1  ~,
  \label{1}\\
  \cos \alpha & = & \frac{k \cdot l}{(u \cdot k) (u \cdot l)} + 1  ~, \\
  \cos \beta & = & \frac{l \cdot w}{(u \cdot l) (u \cdot w)} + 1  ~,
\end{eqnarray}
where $u$ is the observer's 4-velocity and $k \cdot w \equiv g_{\mu \nu} k^{\mu} w^{\nu}$. The schematic diagram is shown in Figure~{\ref{Fig1}},

One can transform the $\alpha, \beta$ and $\gamma$ into celestial coordinates of point $\rm C$,
\begin{eqnarray}
  \Psi & = & \frac{\pi}{2} - \arcsin \left( \sin (\beta) \sqrt{1 - \left(
  \frac{\cos \alpha - \cos \beta \cos \gamma}{\sin \beta \sin \gamma}
  \right)^2} \right)  ~, \\
  \Phi & = & \gamma - \arccos \left( \frac{\cos \beta}{\cos \left(
  \frac{\pi}{2} - \Psi \right)} \right)  ~ .
\end{eqnarray}
The boundary of the shadows is determined by taking all the points of C in celestial coordinates. In order to study the shape of the shadows in two-dimensional plane, stereographic projection is usually used \cite{grenzebach_photon_2014,chang_revisiting_2020},
\begin{eqnarray}
  Y & = & \frac{2 \sin \Phi   \sin \Psi}{1 + \cos \Phi
    \sin \Psi}  ~, \\
  Z & = & \frac{2 \cos \Psi}{1 + \cos \Phi   \sin \Psi}
   ~ .
\end{eqnarray}

Here, we can introduce a distortion parameter $\Xi$ in terms of $\alpha,
\beta$ and $\gamma$,
\begin{equation}
  \cos \Xi \equiv \cos (   \angle  \rm{BCA}) = \frac{1 + \cos \gamma - \cos
  \alpha - \cos \beta}{2 \sqrt{(1 - \cos \alpha) (1 - \cos \beta)}}  ~,
\end{equation}
where $\Xi$ ranges from 0 to $\pi$. The $\cos \Xi \equiv 0$ indicates that shape of shadows is circular in the celestial sphere. For the non-vanished $\cos \Xi$, it can quantify distortion of shape of the shadows from circularity.

\section{Shadow of spherical black hole quantified the distortion
parameters}\label{III}

In this section, we would study spherical black hole shadows with the distortion parameter $\Xi$. We consider the metric of a static spherical black hole in the form of
\begin{equation}
  {\rm d} s^2 = - A   (r) {\rm d} t^2 + B (r) {\rm d} r^2 + r^2 ({\rm d}
  \theta^2 + \sin^2 \theta {\rm d} \phi^2)  ~ .
\end{equation}
By making use of the metric, one can obtain 4-velocities of light rays from null geodesic equations,
\begin{eqnarray}
  \frac{p^0}{E} & = &   \frac{1}{A}  ~,  \label{10}\\
  \frac{p^1}{E} & = & \frac{1}{B} \sqrt{E^2 \left( \frac{B  }{A} -
  \frac{K}{E^2} \frac{B}{r^2} \right)} = \sqrt{\frac{1}{A   B} -
  \frac{\kappa}{B   r^2}}  ~,  \label{11}\\
  \frac{p^2}{E} & = & \frac{1}{E   r^2} \sqrt{K - \frac{L^2}{\sin^2
  \theta}} = \frac{1}{r^2} \sqrt{\kappa - \frac{\lambda^2}{\sin^2 \theta}}
   ~,  \label{12}\\
  \frac{p^3}{E} & = & \frac{L}{E   r^2 \sin^2 \theta} =
  \frac{\lambda}{r^2 \sin^2 \theta}  ~,  \label{13}
\end{eqnarray}
where $E$, $L$ and $K$ are integral constants from null geodesic equations. The $L$ and $K$ have been substituted by $\kappa \equiv \frac{K}{E^2}$ and $\lambda \equiv \frac{L}{E}$, respectively. For the shadow of a black hole, $\kappa$ is determined by  the photon sphere of the black hole, namely, $\left( \frac{{\rm d} r}{{\rm d} \lambda} \right)_{r_{ \rm{ph}}} = 0$ and $\left( \frac{{\rm d}^2 r}{{\rm d} \lambda^2} \right)_{r_{ \rm{ph}}} = 0$. The range of $\lambda$ is determined by $\kappa - \frac{\lambda^2}{\sin^2 \theta} \geqslant 0$. All these lead to
\begin{eqnarray}
  \kappa & = & \frac{r_{ \rm{sp}}^2}{A (r_{ \rm{sp}})}  ~, \\
  \lambda^2 & \leqslant & \frac{r_{ \rm{sp}}^2 \sin^2 \theta}{A
  (r_{ \rm{sp}})}  ~,
\end{eqnarray}
where $r_{ \rm{sp}}$ is determined by solving the equation $r   A' - 2 A
= 0$.

\subsection{Shape of spherical black hole shadows for arbitrary observers}

As the spherical symmetry of the black holes, we could consider observers in the equatorial plane $\theta = \frac{\pi}{2}$ for simplicity. In order to study the shape of the shadows in the view of given observer $u^{\mu} = (u^0, u^1, 0, u^3)$, we can calculate the distortion parameter $\Xi$, directly,
\begin{eqnarray}
  \cos \Xi & = & \frac{1 + \cos \gamma - \cos \alpha - \cos \beta}{2 \sqrt{(1  - \cos \alpha) (1 - \cos \beta)}} \nonumber\\
  & = & \frac{1}{2 \sqrt{(1 - \cos \alpha) (1 - \cos \beta)}} \left( \frac{(k  \cdot l) (u \cdot w) + (w \cdot l) (u \cdot k) - (k \cdot w) (u \cdot l)}{(u  \cdot l) (u \cdot k) (u \cdot w)} \right) \nonumber\\
  & = & \frac{1}{2 (u \cdot l) (u \cdot k) (u \cdot w) \sqrt{(1 - \cos  \alpha) (1 - \cos \beta)}} \left( u_0 \left( (g_{00} k^0 l^0 + g_{11} k^1  l^1 + g_{33} k^3 l^3) w^0 \right. \right. \nonumber\\
  & &  \left. + (g_{00} w^0 l^0 + g_{11} w^1 l^1 + g_{33} w^3  l^3) k^0 - \left( g_{11} k^0 w^0 + g_{11} k^1 w^0 + g_{33} k^3 w^3 \right)  l^0 \right)  \nonumber\\
  & & \left. + u_1 \left( \left( g_{00} k^0 l^0 + g_{11} k^1 l^1 + g_{33} k^3  l^3 \right) w^1  \right. \right. \nonumber\\
  & & \left. \left. + \left( g_{00} w^0 l^0 + g_{11} w^1 l^1 + g_{33} w^3 l^3 \right) k^1 -  \left( g_{11} k^0 w^0 + g_{11} k^1 w^1 + g_{33} k^3 w^3 \right) l^1 \right) \right.   \nonumber \\
  & & \left. + u_3 \left( \left( g_{00} k^0 l^0 + g_{11} k^1 l^1 + g_{33} k^3 l^3 \right)  w^3  \right. \right. \nonumber\\
  & & \left. \left. + \left( g_{00} w^0 l^0 + g_{11} w^1 l^1 + g_{33} w^3 l^3 \right) k^3 -  (g_{00} k^0 w^0 + g_{11} k^1 w^1 + g_{33} k^3 w^3) l^3 \right) \right)
  \nonumber\\
  & = & \frac{u_0 \times 0 + u_1 \times 0 + u_3 \times 0}{2 (u \cdot l) (u  \cdot k) (u \cdot w) \sqrt{(1 - \cos \alpha) (1 - \cos \beta)}} \nonumber\\
  & = & 0  ~,
\end{eqnarray}
where we have inverted the 4-velocities of light rays from
Eqs.~(\ref{11})-(\ref{13}),
\begin{eqnarray}
  k^{\mu} & = &   p^{\mu} |_{\kappa = \frac{r_{ \rm{sp}}^2}{A
  (r_{ \rm{sp}})}, \lambda = \frac{r_{ \rm{sp}}}{\sqrt{A (r_{ \rm{sp}})}}}
   ~, \\
  w^{\mu} & = &   p^{\mu} |_{\kappa = \frac{r_{ \rm{sp}}^2}{A
  (r_{ \rm{sp}})}, \lambda = - \frac{r_{ \rm{sp}}}{\sqrt{A
  (r_{ \rm{sp}})}}}  ~, \\
  l^{\mu} & = &   p^{\mu} |_{\kappa = \frac{r_{ \rm{sp}}^2}{A
  (r_{ \rm{sp}})}, \lambda}  ~.
\end{eqnarray}
It shows that $\Xi \equiv \frac{\pi}{2}$ is independent on the observers' 4-velocities. For spherical black holes, we can conclude that the shape of the shadows of static spherical black holes is circular in the view of arbitrary observers.

\subsection{Size of the shadows for different observers}

\begin{figure}[ht!]
	\centering
	{\includegraphics[width=0.7\linewidth]{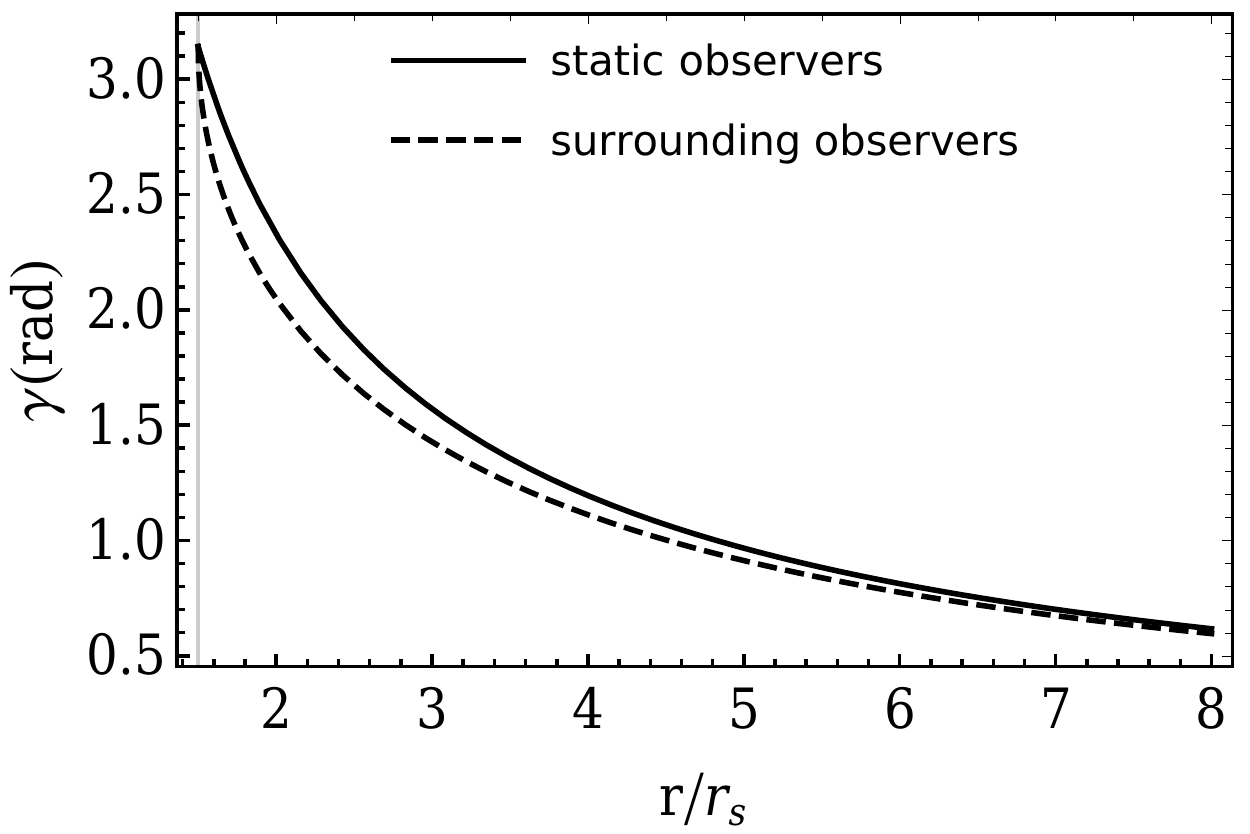}}
	\caption{Angular diameters as a function of the observers' radial coordinate $r$ in Schwarzschild space-time for selected observers. \label{Fig2}}
\end{figure}

Since the shape of the shadows is circular for arbitrary observers, the size of them can be described by the angular diameter $\gamma$ without ambiguity. From Eq.~{}(\ref{1}), one can obtain the relation between angular diameters $\gamma^{(A)}$ and $\gamma^{(B)}$ for different observers $A$ and $B$,
\begin{eqnarray}
  \sin \left( \frac{\gamma^{(A)}}{2} \right) & = & \sin \left(
  \frac{\gamma^{(B)}}{2} \right) \sqrt{\frac{(u^{(B)} \cdot k) (u^{(B)} \cdot
  w)}{(u^{(A)} \cdot k) (u^{(A)} \cdot w)}}  ~ .  \label{20}
\end{eqnarray}
For the sake of intuition, we consider the 4-velocities of the observers $A$
and $B$ in the form of
\begin{eqnarray}
  u^{(B)} & = & - \sqrt{A} {\rm d} t  ~, \\
  u^{(A)} & = & - \sqrt{\frac{r^3 \partial_1 \ln \sqrt{A}}{1 - r \partial_1
  \ln \sqrt{A}}} {\rm d} t + \sqrt{\frac{A}{1 - r^2 \partial_1 \ln \sqrt{A}}}
  {\rm d} \phi  ~ .
\end{eqnarray}
They are static and geodesically surrounding with respect to the centre black hole,
respectively. We can rewrite Eq.~(\ref{20}) as
\begin{eqnarray}
  \sin \left( \frac{\gamma^{(A)}}{2} \right) & = & \sin \left(
  \frac{\gamma^{(B)}}{2} \right) \sqrt{\frac{1 - r \partial_1 \ln \sqrt{A}}{1
  - \frac{A \kappa}{r} \partial_1 \ln \sqrt{A}}}  ~ .
\end{eqnarray}
In Figure~{\ref{Fig2}}, we plot the angular diameters as a function of observers' radial coordinate $r$ in Schwarzschild space-time. One may find that the motion of observers would lead to a decrease of the size of the shadows. As known that the shape of the shadows is circular, the diameter of the shadows in the direction perpendicular to the observers' motion should also decrease. It is noticed that this is hard to be understood as the length contraction effect in special relativity.

\begin{figure}[ht]
	\centering
	{\includegraphics[width=0.7\linewidth]{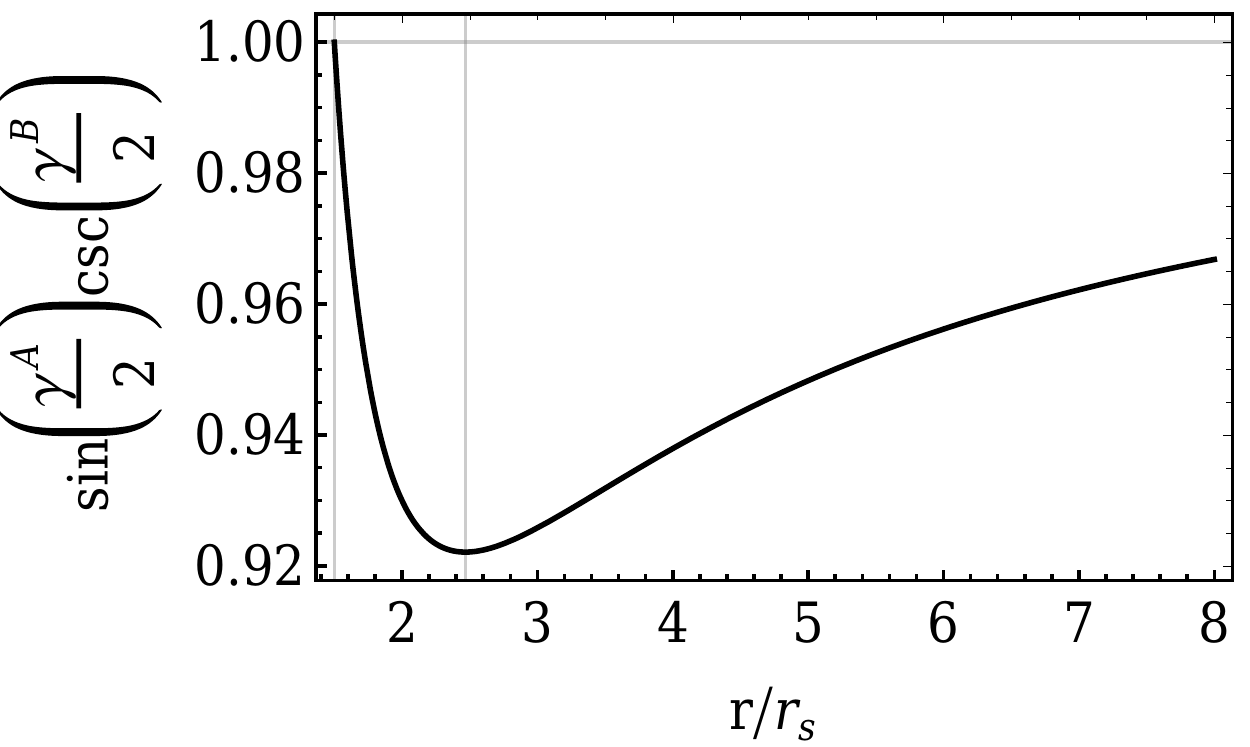}}
	\caption{Ratio $\sin (\gamma^{(A)} / 2) / \sin (\gamma^{(B)} / 2)$ as function of observers' radial coordinate $r$ in Schwarzschild space-time for observers A and B. \label{Fig3}}
\end{figure}

In Figure~{\ref{Fig3}}, we present the ratio $\sin (\gamma^{(A)} / 2) / \sin (\gamma^{(B)} / 2)$ as a function of the observers' radial coordinate $r$ in Schwarzschild space-time. It shows that the ratio is not monotonic. The extreme point of
it is
\begin{equation}
  r_{ \rm{ex}} = \frac{3 r_s}{2} \left( \sqrt{7} - 1 \right)  ~ ,
\end{equation}
where $r_s$ is Schwarzschild radius. It suggests that the $r_{ \rm{ex}}$ might be used to divide the near region and the far region of a black hole. In the far region, the propagation effect of light would dominate. As the speed of the observers increases, contraction of the size of the shadows in the view of these observers also increases. In the near region, the propagation effect of light turns out to be not important.  In this region, the space-time tends to be absolute.

\section{Kerr black hole shadows for surrounding observers}\label{IV}

We have shown that the shape of a spherical black shadow is independent on motional status of observers. However, as suggested in Ref.~\cite{chang_revisiting_2020}, it seems not true for rotating black hole shadows. In this section, we turn to rotating black hole shadows in the view of surrounding observers by making use of the astrometrical observable approach. Here, for simplicity, we consider Kerr black holes as an instance,
\begin{equation}
  {\rm d} s^2 = - \frac{\Delta}{\Sigma} ({\rm d} t -  a \sin^2 \theta {\rm d}
  \phi)^2 + \frac{\sin^2 \theta}{\Sigma} (a {\rm d} t - (r^2 + a^2) {\rm d}
  \phi)^2 + \frac{\Sigma}{\Delta} {\rm d} r^2 + \Sigma {\rm d} \theta^2  ~,
\end{equation}
where
\begin{eqnarray}
  \Delta & = & r^2 - 2 M   r + a^2  ~, \\
  \Sigma & = & r^2 + a^2 \cos^2 \theta  ~ .
\end{eqnarray}
From null geodesic equations, the 4-velocities of light rays take the form of
\begin{eqnarray}
  \Sigma p^0 & = & E \left( \frac{(r^2 + a^2 - a \lambda) (r^2 + a^2)}{\Delta}
  + a (\lambda -  \rm{asin}^2 \theta) \right)  ~, \\
  (\Sigma p^1)^2 & = & R (r)  ~, \\
  (\Sigma p^2)^2 & = & \Theta (\theta)  ~, \\
  \Sigma p^3 & = & E \left( \frac{a (r^2 + a^2) - a^2 \lambda}{\Delta} +
  \frac{\lambda -  \rm{asin}^2 \theta}{\sin^2 \theta} \right)  ~,
\end{eqnarray}
where
\begin{eqnarray}
  R (r) & \equiv & E^2 ((r^2 + a^2 - a \lambda)^2 - \Delta \kappa)  ~, \\
  \Theta (\theta) & \equiv & E^2 \left( \kappa - \frac{(\lambda -
   \rm{asin}^2 \theta)^2}{\sin^2 \theta} \right)  ~,
\end{eqnarray}
and
\begin{eqnarray}
  \lambda & \equiv & \frac{L}{E}  ~, \\
  \kappa & \equiv & \frac{K}{E^2}  ~ .
\end{eqnarray}
The integral constants $\lambda$ and $\kappa$ are determined by the photon region
of Kerr black holes,
\begin{eqnarray}
  \lambda (r_c) & = & \frac{1}{a} \left( r^2 + a^2 - \frac{4 r
  \Delta}{\Delta'} \right)_{r = r_c}  ~, \\
  \kappa (r_c) & = & \left( \frac{16 r^2 \Delta}{(\Delta')^2} \right)_{r =
  r_c}  ~ .
\end{eqnarray}
The range of $r_c$ is determined by $\Theta (\theta) \geqslant 0$, namely,
\begin{equation}
  ((4 r \Delta_r - \Sigma \Delta_r')^2 - 16 a^2 r^2 \Delta_r \sin^2 \theta)_{r
  = r_c} \leqslant 0  ~ .
\end{equation}
 These results have been studied carefully in preview works \cite{bardeen_timelike_1973,grenzebach_photon_2014}.

\subsection{Shadow in the view of surrounding observers}

The most interesting and simplest surrounding observers are those located at equatorial plane $\theta = \frac{\pi}{2}$. Here, more specifically, we consider the shadows in the view of geodesic observers, ZAMOs and Carter observers as representative cases.

\subsubsection{Geodesic observers}

From geodesic equations and $\frac{{\rm d}^2 r}{{\rm d} \tau^2} = \frac{{\rm d} r}{{\rm d} \tau} = 0$, we have
\begin{equation}
  \Omega_{\pm} \equiv  \frac{u^3}{u^0} = \mp
  \sqrt{\frac{M}{r^3}} + \frac{a   M}{a^2 M - r^3} \left( \frac{1}{2}
  \pm \sqrt{\frac{a^2 M}{r^3}} \right)  ~,
\end{equation}
where $\pm$ represents anticlockwise and clockwise motion of the geodesic observers. Associated $u^{\mu} u_{\mu} = - 1$, the 4-velocities take the form of
\begin{equation}
  u_{ \rm{geo}, \pm} = \frac{1}{\sqrt{- g_{00} - 2 g_{03} \Omega_{\pm} -
  g_{33} \Omega^2}} \partial_0 + \frac{\Omega_{\pm}}{\sqrt{- g_{00} - 2 g_{03}
  \Omega_{\pm} - g_{33} \Omega^2_{\pm}}} \partial_3  ~ .
\end{equation}

\begin{figure}[ht]
	\centering
	{\includegraphics[width=1\linewidth]{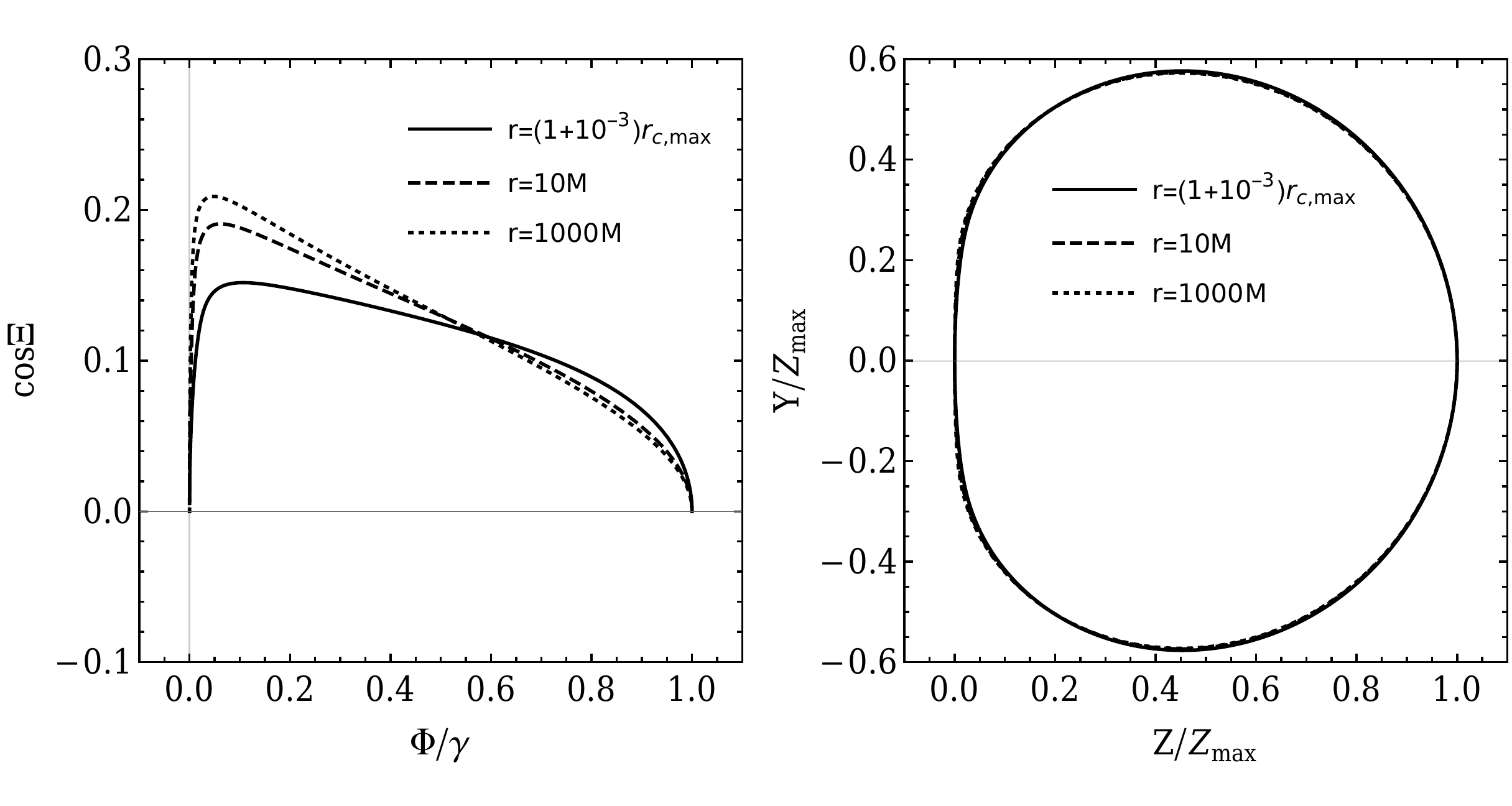}}
	\caption{Left panel: distortion parameter $\Xi$ of the shadows as a function of $\frac{\Phi}{\gamma}$ for observers $u_{ \rm{geo}, -}$ at selected position $r$. Right panel: the corresponding shape of shadows in the left panel. Here, the spin parameter $a=0.999$.\label{Fig4}}
\end{figure}

In Figure~\ref{Fig4}, we present the $\Xi$ as a function of $\frac{\Phi}{\gamma}$ for observers $u_{ \rm{geo}, -}$ at selected position $r$ and corresponding shape of the shadows.
It shows that the distortion of shadow would increase with the observers' radial coordinate $r$. The shape of shadows seems very closed to each other. Without the distortion parameter $\Xi$, it seems difficult to quantify the difference of the shape of these shadows.

\begin{figure}[ht]
	\centering
	{\includegraphics[width=1\linewidth]{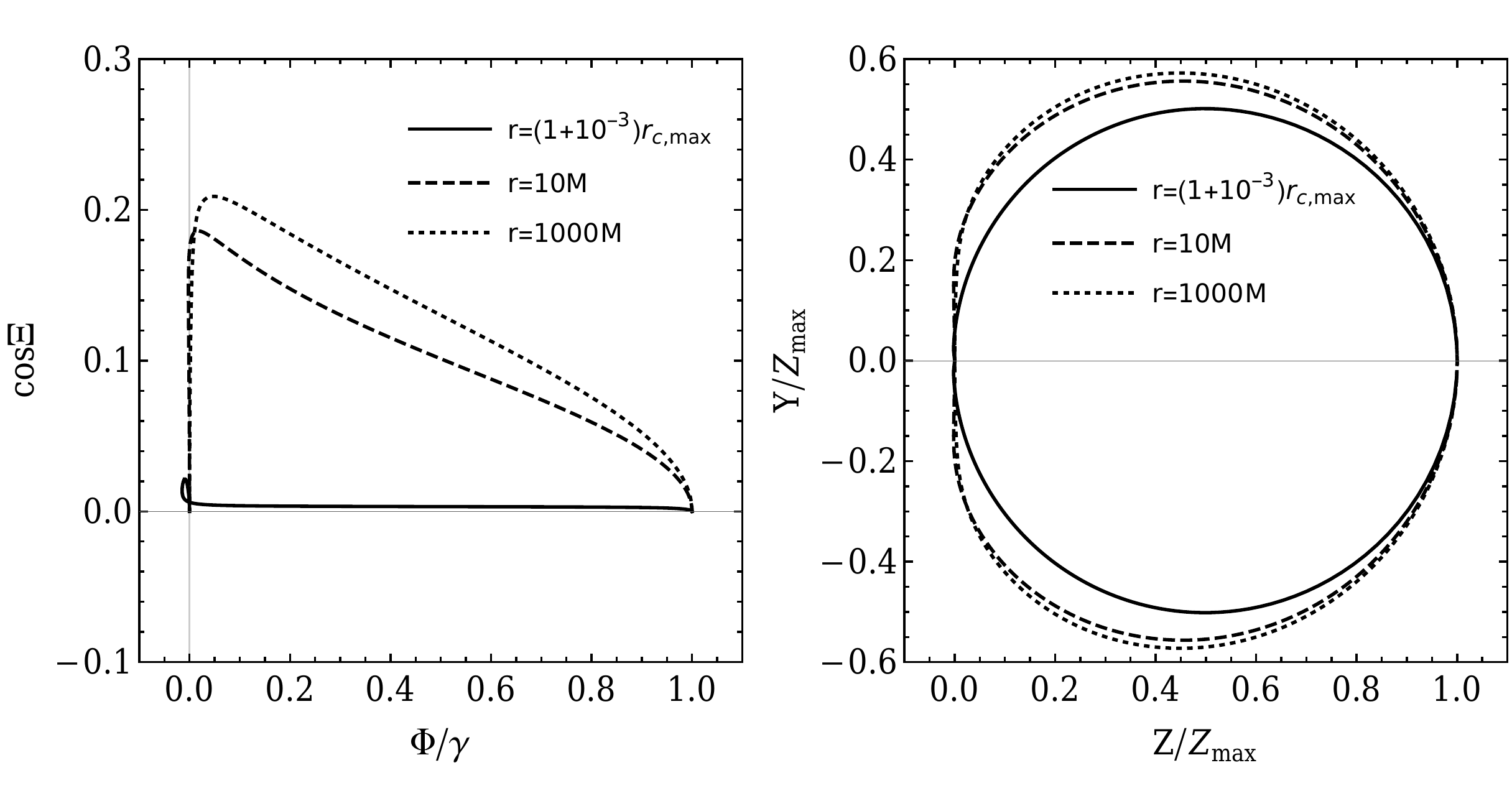}}
	\caption{Left panel: distortion parameter $\Xi$ of the shadows as a function of $\frac{\Phi}{\gamma}$ for observers $u_{ \rm{geo}, +}$ at selected position $r$. Right panel: the corresponding shape of shadows in the left panel. Here, the spin parameter $a=0.999$. \label{Fig5}}
\end{figure}

We also present the $\Xi$ as a function of $\frac{\Phi}{\gamma}$ in the view of observers $u_{ \rm{geo}, +}$ in Figure~\ref{Fig5}. It is interesting to find that the shape of the shadows tends to be circular when the surrounding observers are closed to the Kerr black holes. It provides a picture that the observers might not find distortion of the shadows from circularity, if they are co-rotating with the Kerr black hole, closely.

\subsubsection{Zero angular momentum observers}

In the first paper on rotating black shadows, Bardeen calculated the Kerr black hole shadows with orthogonal tetrads in the view of ZAMOs,
\begin{equation}
  u_{ \rm{ZAMO}} = \frac{1}{r} \sqrt{\frac{(r^2 + a^2)^2 - a^2
  \Delta}{\Delta}} \partial_0 + \frac{2 a   M  }{\sqrt{\Delta
  ((r^2 + a^2)^2 - a^2 \Delta)}} \partial_3  ~ .
\end{equation}
Here, we alternatively calculate the Kerr black hole shadows with astrometrical observables in the view of ZAMOs.

\begin{figure}[ht]
	\centering
	{\includegraphics[width=1\linewidth]{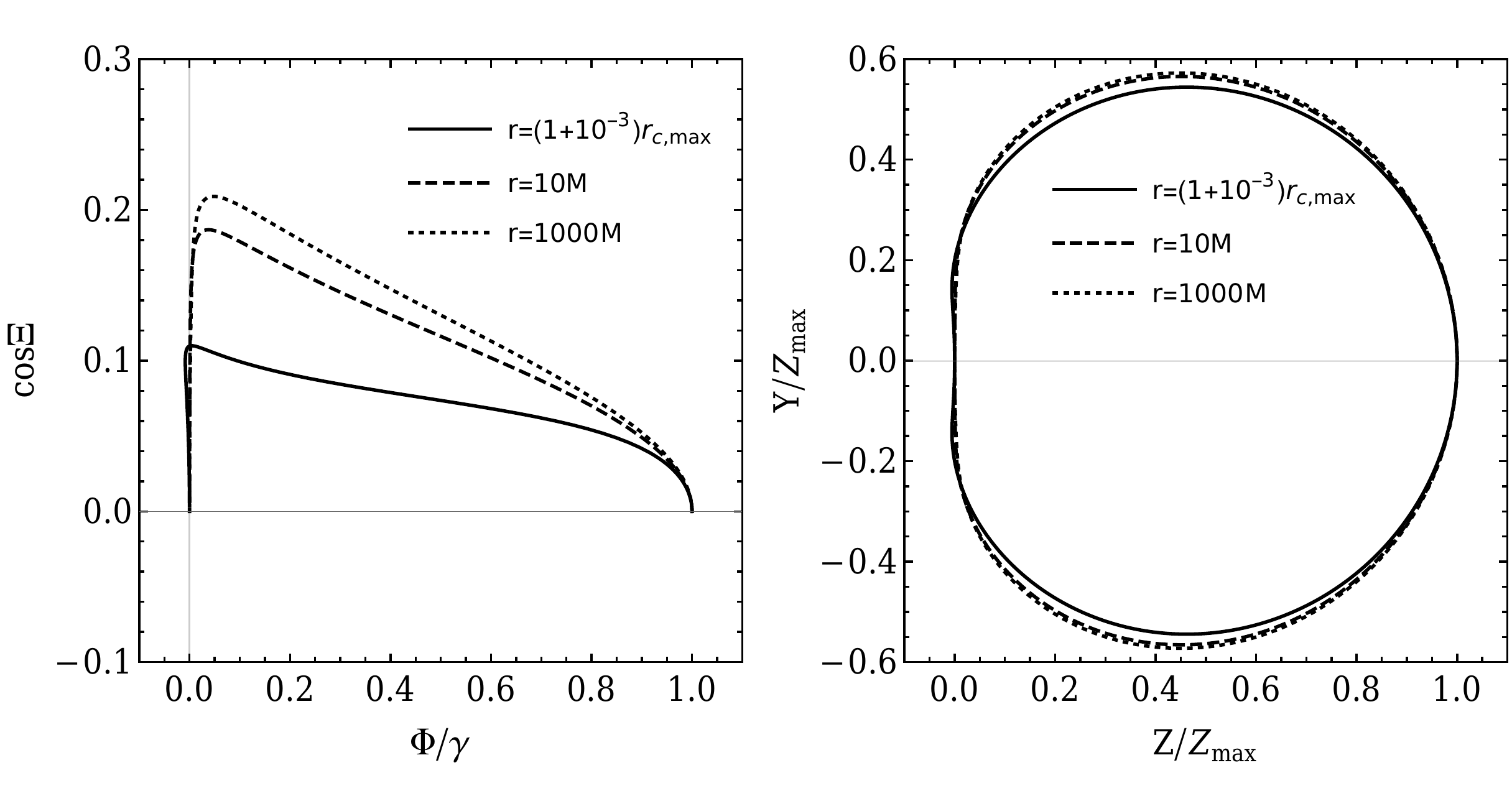}}
	\caption{Left panel: distortion parameter $\Xi$ of the shadows as function of $\frac{\Phi}{\gamma}$ for observers $u_{ \rm{ZAMO}}$ at selected position $r$. Right panel: the corresponding shape of shadows in the  left panel. Here, the spin parameter $a=0.999$. \label{Fig6}}
\end{figure}

In Figure~\ref{Fig6}, the $\Xi$ as  a function of $\Phi /\gamma$ for selected observers and corresponding shape of the shadows are presented.
It shows that the distortion of the shadows increases with radial coordinate $r$. The results seem different from that obtained by Bardeen \cite{bardeen_timelike_1973,chang_revisiting_2020}. We would further compare Bardeen's approach with the astrometrical observable approach in the next section.

\subsubsection{Carter's observers}

Also, we consider Carter observers that take the form of
\begin{equation}
  u_{ \rm{car}} = \frac{r^2 + a^2}{r \sqrt{\Delta}} \partial_0 + \frac{a}{r
  \sqrt{\Delta}} \partial_3  ~ .
\end{equation}
This type of observers is first used by Grenzabach for calculating Kerr-like black hole shadows in orthogonal tetrads. Here, we alternatively calculate the Kerr black hole shadows with astrometrical observables in the view of Carter's observers.

\begin{figure}[ht]
	\centering
	{\includegraphics[width=1\linewidth]{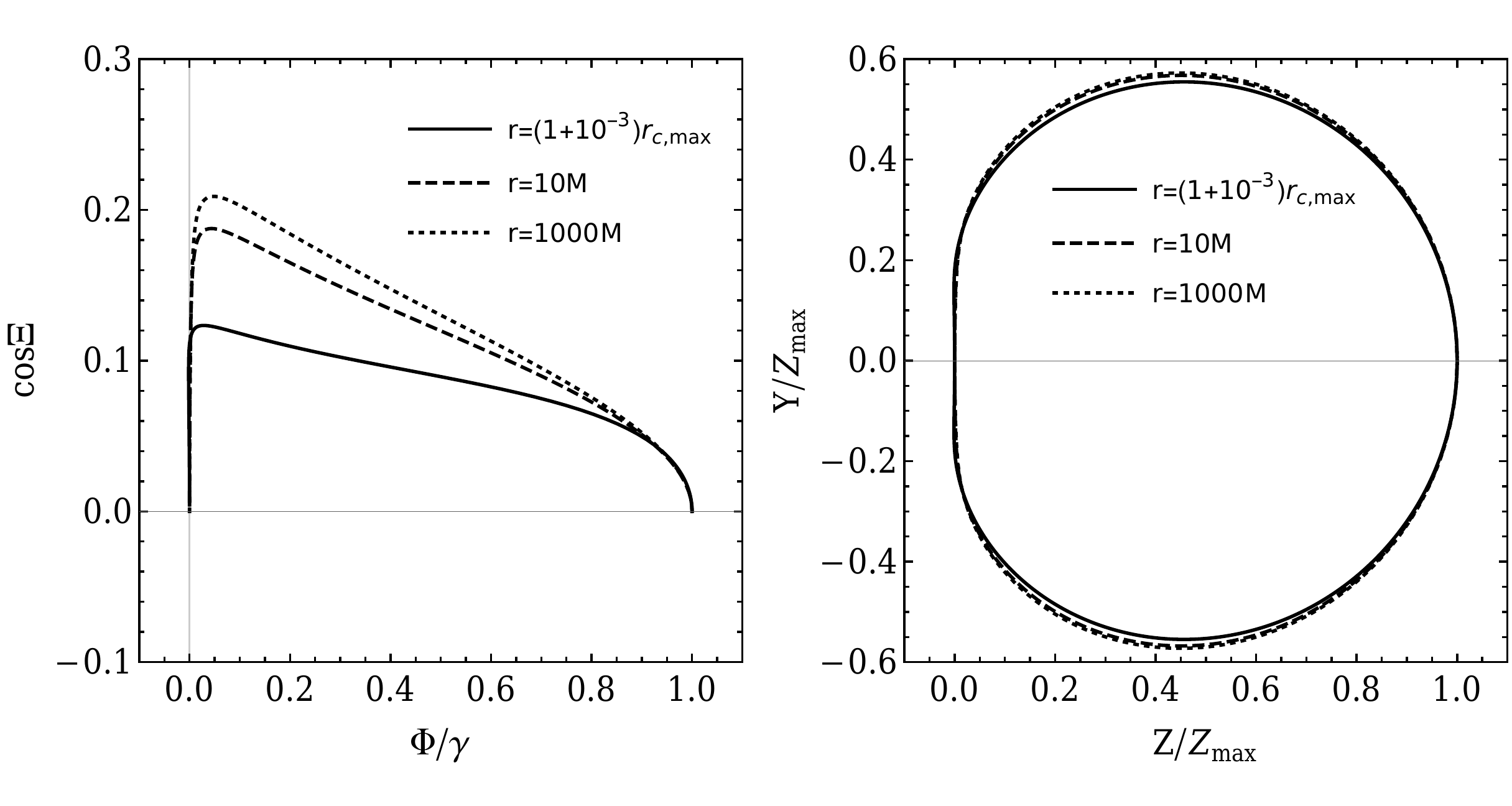}}
	\caption{Left panel: distortion parameter $\Xi$ of the shadows as a function of $\frac{\Phi}{\gamma}$ for observers $u_{ \rm{Car}}$ at selected position $r$. Right panel: the corresponding shape of shadows in the left panel. Here, the spin parameter $a=0.999$. \label{Fig7}}
\end{figure}

In Figure~\ref{Fig7}, we present the $\Xi$ as a function of $\Phi / \gamma$ for the selected observers and corresponding shape of the shadows.
It also shows that the distortion of the shadows increases with observers' radial coordinate $r$.

\subsubsection{Comparisons}

\begin{figure}[ht]
	\centering
	{\includegraphics[width=1\linewidth]{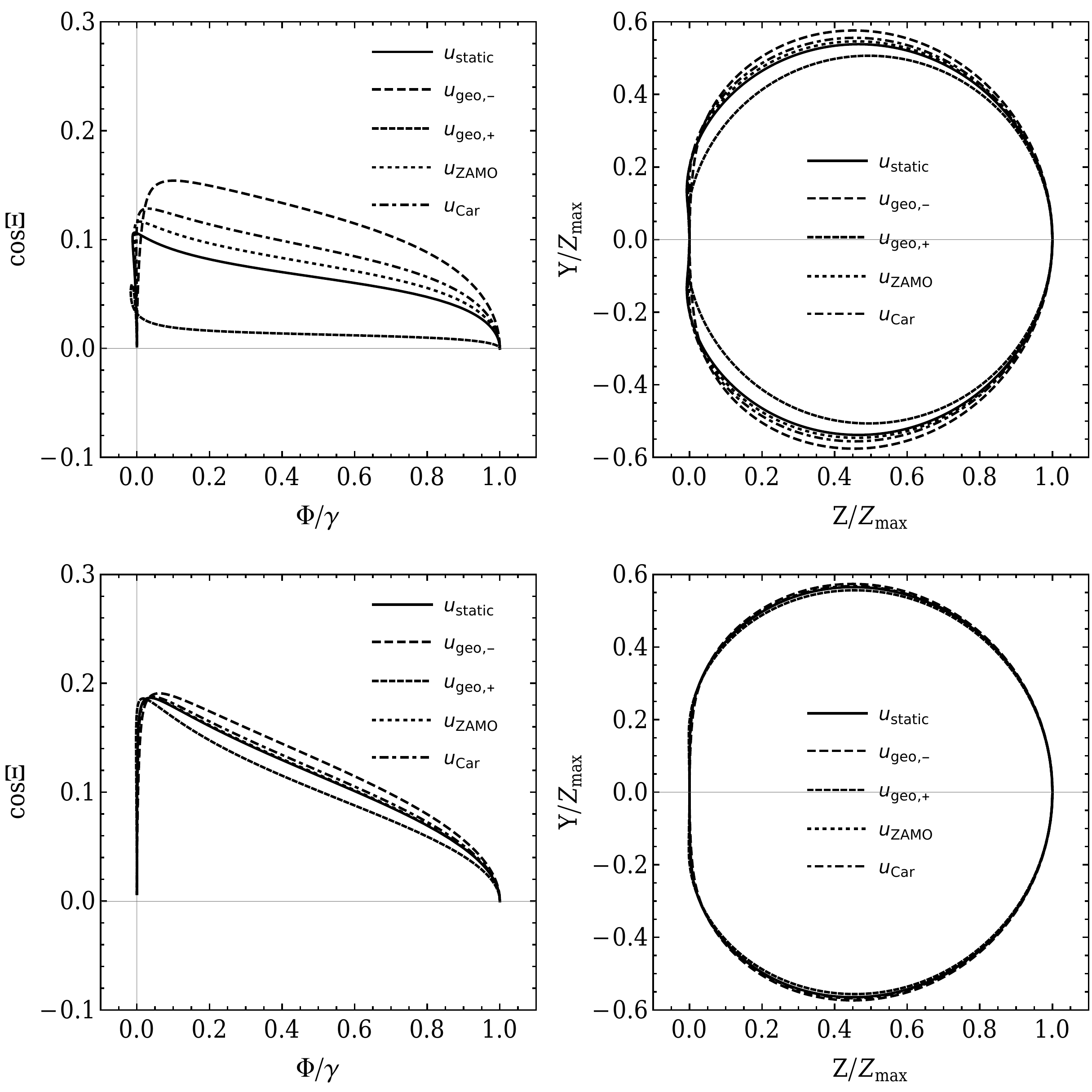}}
	\caption{Top-left panel: distortion parameter $\Xi$ of the shadows as a function of $\frac{\Phi}{\gamma}$ for selected observers at position $r=4$M. Top-right panel: the corresponding shape of shadows in top-left panel.
	Bottom-left panel: distortion parameter $\Xi$ of the shadows as a function of $\frac{\Phi}{\gamma}$ for selected observers at position $r=14$M. Bottom-right panel: the corresponding shape of shadows in the bottom-left panel. Here, the spin parameter $a=0.999$. \label{Fig8}}
\end{figure}

In order to compare all of the types of the observers, we present the $\Xi$ as a function of $\frac{\Phi}{\gamma}$ for these observers at position $r = 5 M$ and 14$M$ in Figure~\ref{Fig8}.
For all of the types of the observers $\frac{\Phi}{\gamma} \lesssim 0.1$, the shapes of shadows are highly deviated from circularity. For all of the types of the observers in the far region, the shape of the shadows tends to be the same with each other. The distortion of shadows in view of static observers, ZAMOs and Carter observers are between those in view of geodesic observers $u_{ \rm{geo}, +}$ and $u_{ \rm{geo}, -}$. \ We can conclude that the distortion of Kerr black hole shadows is dependent on the motion status of observers, especially for those in the near region. It is different from spherical black holes.

\subsection{Is there a surrounding observer that could view the shape of Kerr black hole shadows as circularity? }

As the shapes of Kerr black hole shadows are dependent on the motional status of observers, one might question if there is a special type of surrounding observers that could view the shape of Kerr black hole shadows as circularity. In order to respond to this problem, we have to look for a solution of equation $\cos \Xi \equiv 0$ for surrounding observers $u = (u^0, 0, 0, u^3)$,
\begin{eqnarray}
  \cos \Xi & = & \frac{1 + \cos \gamma - \cos \alpha - \cos \beta}{2 \sqrt{(1
  - \cos \alpha) (1 - \cos \beta)}} \nonumber\\
  & = & \frac{1}{2 (u \cdot l) (u \cdot k) (u \cdot w) \sqrt{(1 - \cos
  \alpha) (1 - \cos \beta)}} \left( u_0 \left( l^0 (- g_{00} k^0 w^0 + g_{11}
  k^1 w^1 + g_{33} k^3 w^3)  \right. \right. \nonumber\\
  & & \left. \left.  + l^1 \left( - g_{11} k^1 w^0 - g_{11} k^0 w^1
  \right) + l^3 (- g_{33} k^3 w^0 - g_{33} k^0 w^3 - 2 g_{03} k^0 w^0) \right) \right. \nonumber \\
  & & \left.
  + u_3 \left( l  ^0 (- g_{00} k^0 w^3 - g_{00} k^3 w^0 - 2 g_{03} k^3
  w^3)  \right. \right. \nonumber\\
  & & \left. \left. + l^1 \left( - g_{11} k^1 w^3 - g_{11} k^3 w^1 \right) + l^3 (g_{00}
  k^0 w^0 + g_{11} k^1 w^1 - g_{33} k^3 w^3) \right) \right)  ~,
\end{eqnarray}
where
\begin{eqnarray}
  k^{\mu} & = &   p^{\mu} |_{\kappa (r_{c, \min}), \lambda (r_{c,
  \min})}  ~, \\
  w^{\mu} & = &   p^{\mu} |_{\kappa (r_{c, \max}), \lambda (r_{c,
  \max})}  ~, \\
  l^{\mu} & = &   p^{\mu} |_{\kappa (r_c), \lambda (r_c)}  ~ .
\end{eqnarray}
As $\cos \Xi \equiv 0$ should be independent on $r_c$, we can find a solution in the form of
\begin{equation}
  \frac{u_3}{u_0} = - \frac{k^1 w^0 + k^0 w^1}{k^1 w^3 + k^3 w^1}  ~ .
\end{equation}
It can be rewritten as
\begin{equation}
  \frac{u^3}{u^0} = - \frac{w_0 k^1 + k_0 w^1}{w_3 k^1 + k_3 w^1} = - \frac{1
  + \sqrt{\frac{R(r)|_{r{_c,\rm max}}}{R(r)|_{r_{c,\rm min}}}}}{\lambda(r_{c,\rm min}) + \lambda(r_{c,\rm max}) \sqrt{\frac{R(r)|_{r_{c,\rm max}}}{R(r)|_{r_{c,\rm min}}}}}  ~
  .
\end{equation}
There seems to be a type of surrounding observers so as to $\cos \Xi \equiv 0$. However, if considering the inner product of the 4-velocities, we have
\begin{eqnarray}
  u \cdot u & = & g_{00} (u^0)^2 + g_{33} (u^3)^2 + 2 g_{03} (u^3)^2
  \nonumber\\
  & = & (u^0)^2 \left( g_{00} + g_{33} \left( \frac{u^3}{u^0} \right)^2 + 2
  g_{03} \left( \frac{u^3}{u^0} \right) \right) > 0  ~,
\end{eqnarray}
It indicates that this type of observers is space-like or super-luminal. Thus, there is not a physical surrounding observer that could view the shape of the Kerr black hole shadow as circularity. It might suggest that it is difficult for a telescope to ignore spin of a Kerr black hole. In the point of view  of black hole shadows, the Schwarzschild and Kerr black holes are shown to be very different.

\section{Comparison with orthogonal tetrad approaches\label{V}}

In section~\ref{IV}, we have calculated the Kerr black hole shadow with astrometrical observables for given observers. As suggested in Ref.~\cite{chang_revisiting_2020}, this
approach could be different from orthogonal tetrad approaches \cite{bardeen_timelike_1973,grenzebach_photon_2014}.

In order to compare these approaches, we use a distortion parameter for the shadows in 2D plane  \cite{hioki_measurement_2009,chang_revisiting_2020},
\begin{equation}
  \delta \equiv 1 - \frac{D_{\min}}{D_{\max}}  ~ .
\end{equation}

\begin{figure}[ht]
	\centering
	{\includegraphics[width=0.7\linewidth]{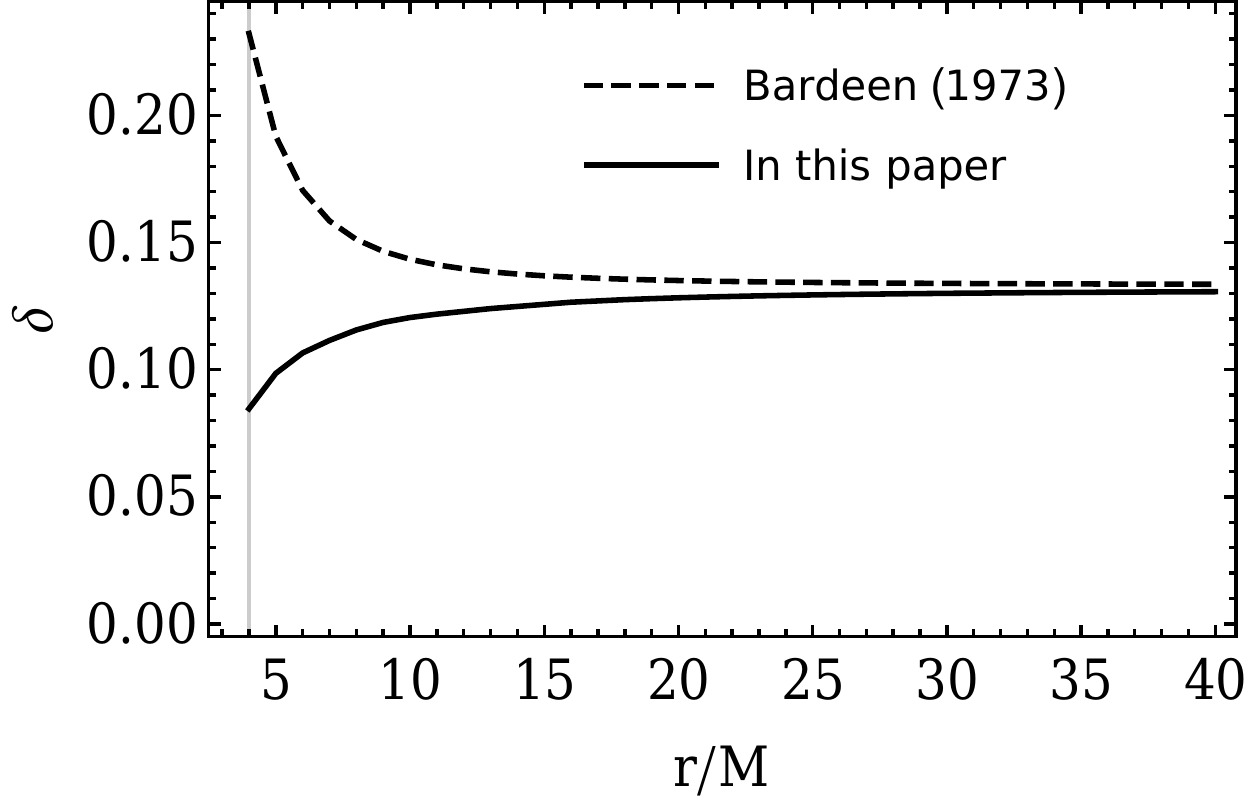}}
	\caption{distortion parameter $\delta$ of the Kerr black hole shadows as a function of ZAMOs' radial coordinate $r$ using Bardeen's approach and the astrometrical observable approach. Here, we choose the spin parameter $a=0.9999$. \label{Fig9}}
\end{figure}

In the pioneer work of Bardeen \cite{bardeen_timelike_1973}, the Kerr black hole shadow was calculated with respect to ZAMOs in orthogonal tetrads. We can alternatively calculate the shadow in the view of ZAMOs with astrometrical observables. In Figure~\ref{Fig9}, we present the distortion parameter $\delta$ of the shadows as a function of ZAMOs' radial coordinate $r$ using these two approaches.
It shows that the distortion parameter $\delta$ increases with the observers' radical coordinate $r$ in astrometric observables approaches. It is contracted with Bardeen's result. We might think the difference raised from that Bardeen's approach is not equipped with stereographic projection.

\begin{figure}[ht]
	\centering
	{\includegraphics[width=0.7\linewidth]{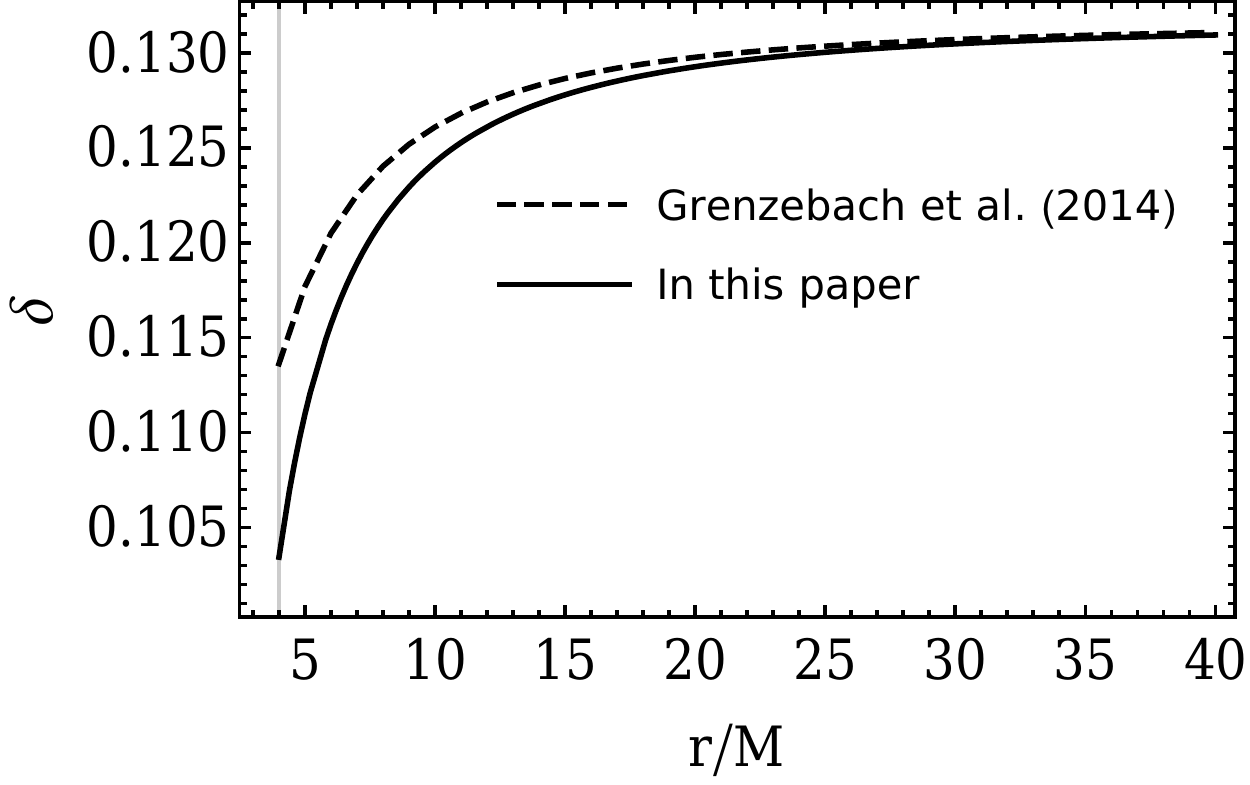}}
	\caption{distortion parameter $\delta$ of the Kerr black hole shadows as  a function of ZAMOs' radial coordinate $r$ using the approach by Grenzabach $et~al.$ and the astrometrical observable approach. Here, we choose the spin parameter $a=0.9999$. \label{Fig10}}
\end{figure}

In the work of Grenzebach $et~al.$ \cite{grenzebach_photon_2014}, the Kerr black hole shadow was calculated with respect to Carter's observers in orthogonal tetrads. In their approach, the stereographic projection is used. Thus, it would be fair to compare the astrometrical observables approaches with theirs. In Figure~\ref{Fig10}, we present the distortion parameter $\delta$ of the shadows as function of Carter observers' radial coordinate $r$ using these two approaches.
It shows that the distortion parameter increases with observers' radical coordinate $r$ in both approaches. Although the results in different approaches are closed to each other, in principle, they are different.

\section{Conclusions and discussions}
In this paper,  we utilized the approach for calculating size and shape of the black hole shadows in terms of astrometrical observables. The size and shape of the shadow of static spherical black holes and Kerr black holes were discussed in a uniform framework. In order to study the shape of the shadows, we introduced a distortion parameter that can quantify the distortion of shadow from circularity.  We showed that
  the shape of the shadow of a spherical black hole is circular in the view of arbitrary observers. For size of the shadows, it tends to be shrunk in the view of a moving observer. In this situation, the diameter of the shadows is contracted even in the direction perpendicular to the observers' motion. This seems not to be understood as length contraction effect in special relativity. On the other side, the shape of Kerr black holes is shown to be dependent on motional status of observers located at finite distance. In spite of this, it is found that there is not a surrounding observer that could view the shape of the Kerr black hole shadows as circularity.
  As the space-time geometry with respect to this type of observers could be simple, we might expect that there could be the observers in other rotating black holes.
  In the final part of this paper, we also compared our approach with preview works.
  We could conclude that the orthogonal tetrads approaches \cite{bardeen_timelike_1973,grenzebach_photon_2014} and astrometrical observable approach for the shadows in the view of finite-distance observers \cite{chang_revisiting_2020} are not consistent with each other.    
 
Besides, we might expect that these results could be helpful for observation of the Sagittarius A* in the centre of the Milky Way, as our solar system is moving surround the centre black hole.

\acknowledgments{
	This work has been funded by the National Nature Science Foundation of China under grant No. 11675182 and 11690022.
}

\bibliography{citation}
\end{document}